\documentclass[12pt]{article}
\usepackage{amssymb,amsmath,amsthm}

\newtheorem{proposition}{Proposition}
\begin{document}

\title{Euclidean Clifford Algebra}
\author{V. V. Fern\'{a}ndez$^{1}\thanks{%
e-mail: vvf@ime.unicamp.br}$, A. M. Moya$^{1}\thanks{%
e-mail: moya@ime.unicamp.br}$ and W. A. Rodrigues Jr.$^{2}\thanks{%
e-mail: walrod@ime.unicamp.br}$ \\
$\hspace{-0.5cm}^{1}$ Institute of Mathematics, Statistics and Scientific
Computation\\
IMECC-UNICAMP CP 6065\\
13083-970 Campinas-SP, Brazil\\
$^{2}$ Department of Mathematics, University of Liverpool\\
Liverpool, L69 3BX, UK}
\date{10/30/2001}
\maketitle

\begin{abstract}
Let $V$ be a $n$-dimensional real vector space. In this paper we introduce
the concept of \emph{euclidean }Clifford algebra $\mathcal{C\ell }(V,G_{E})$
for a given euclidean structure on $V$, i.e., a pair $(V,G_{E})$ where $%
G_{E} $ is a euclidean metric for $V$ (also called an euclidean scalar
product). Our construction of $\mathcal{C\ell }(V,G_{E})$ has been designed
to produce a powerful computational tool. We start introducing the concept
of \emph{multivectors} over $V$. These objects are elements of a linear
space over the real field, denoted by $\bigwedge V.$ We introduce moreover,
the concepts of exterior and euclidean scalar product of multivectors. This
permits the introduction of two \emph{contraction operators }on $\bigwedge
V, $ and the concept of euclidean \emph{interior} algebras. Equipped with
these notions an euclidean Clifford product is easily introduced. We worked
out with considerable details several important identities and useful
formulas, to help the reader to develope a skill on the subject, preparing
himself for the reading of the following papers in this series.
\end{abstract}

\tableofcontents

\section{Introduction}

This is the first paper of a series of seven. We introduce $\mathcal{C\ell }%
(V,G_{E})$, an \emph{euclidean} Clifford algebra of multivectors associated
to an euclidean structure on a $n$-dimensional real vector space $V$. By an
euclidean structure we mean a pair $(V,G_{E})$ where $G_{E}$ is an euclidean
metric on $V.$ Our construction of $\mathcal{C\ell }(V,G_{E})$ has been
designed in order to produce a powerful computational tool. It starts by
introducing the concept of multivectors over $V$. These objects are elements
of a linear space over the real field, denoted by $\bigwedge V.$ We
introduce in $\bigwedge V$, the concepts of \emph{exterior product} and
\emph{euclidean} \emph{scalar product} of multivectors. This permits the
introduction of two \emph{contraction operators} on $\bigwedge V,$ and the
concept of euclidean \emph{interior} algebras. Equipped with these notions
an euclidean Clifford product is introduced. We worked out with considerable
details several important identities and useful formulas, to help the reader
to develope a skill on the subject, preparing himself for the reading of the
following papers in this series\footnote{%
The papers in this series will be denoted when quoted within or in another
paper in the series by I, II, III,...} (and also for the ones in two
forthcoming series of papers). We have the following resume concerning to
the content of the other papers of the present series.

In paper II we introduce the fresh concept of \emph{general} \emph{extensors.%
} The theory of these objects is developed and the properties of some
particular extensors that appear frequently in our theory of multivector
functions and multivector functionals (the subject of papers V,VI and VII)
are worked in details.

In paper III we study the relationship between the concepts of a \emph{%
metric tensor} $G$ (of arbitrary signature $s=p-q$ ( or $(p,q)$ as
physicists say), with $p+q=n)$ for a $n$-dimensional vector space $V$ and a
\emph{metric extensor} $g$ for that space (with the same signature), showing
the advantage of using the later object even in elementary linear algebra
theory. It is worthwhile to emphasize here that our introduction of the
concept of metric extensor plays a crucial role in our definition of\emph{\
metric} Clifford algebras, which are introduced in paper IV. There, a \emph{%
metric} Clifford product is introduced by a well-defined \emph{deformation}
(induced by $g$) of a given euclidean Clifford product. We introduce also
the concept of \emph{gauge metric} \emph{extensor} $h$ associated to a
metric extensor $g,$ and present and prove the so-called \emph{golden} \emph{%
formula.} The gauge extensors\footnote{%
More precisely, in differential geometry the key objects are gauge extensor
\emph{fields}.} appear naturally in our theory of the differential geometry
on manifolds.

In papers V and VI we present a theory of \emph{multivector functions} of a
real variable, and a theory of multivector functions of a $p$-vector
variable. The notions of limit, continuity and differentiability are
carefully studied. Particular emphasis is given to relate the basic concept
of \emph{directional derivative} with other types of derivatives such as the
\emph{generalized} curl, divergence and gradient.

In paper VII we develop a theory of \emph{multivector functionals}, a key
concept for the developments that we have in mind, and that has not been
properly studied until now.

In two future series of papers the material developed in the present series
will be used as the expression and calculational tool of several different
mathematical and physical theories. We quote here our theory of possible
different kinds of \emph{covariant derivatives operators} for \emph{%
multivector} and \emph{extensor\ fields }on arbitrary metric manifolds which
explicitly shows how these different possible covariant derivative operators
are related through the concept of gauge extensors. In particular, the
concept of \emph{deformed} derivative operators will be seen to play a key
rule in our formulation of families of mathematically possible geometric
theories of gravitation (using Clifford algebras methods). We believe that
our presentation of these theories have a new flavor in relation to old
presentations of possible theories of gravitation (of the
Riemann-Cartan-Weyl types). Also, it will be seen that the concept of
multivector functionals (developed in paper VII) is a necessary one for a
presentation of a rigorous formulation of a Lagrangian theory for
multivector and extensor fields having support on an arbitrary manifold (or
subsets of it) representing spacetime.\footnote{%
The definition of a spacetime will be given at the appropriate place.}
\footnote{%
Our approach immediately suggests possible improvements of the Einstein's
gravitational theory as well as other interpretations, where the
gravitational field, contrary to what happens in Einstein's theory is
understood as a physical field in the sense of Faraday.}

Before starting our enterprise we recall that Clifford Algebras and their
applications in Mathematical Physics are now respectable subjects of
research whose wealth can be appreciate by an examination of the topics
presented in the last five international conference on this subject\footnote{%
The subject has even a journal, \emph{Advances in Applied Clifford Algebras}
, edited by J. Keller and in publication since 1991. Keller is an
enthusiastic of the applications of Clifford algebras in theoretical physics
and contributed with several beautiful papers to the subject. His ideas are
nicely described in his recent book \cite{19}.} (\cite{1},\cite{6}). We have
no intention to present here even a small history of the subject, and we do
not claim even to have given any reasonable list of references on papers and
books on the subject. Only a few points and references will be recalled here%
\footnote{%
We antecipately apologize to the author of any important contribution on the
subject that has not been quoted in our brief account.}. Clifford algebras%
\footnote{%
Formulations of Maxwell and Dirac theories which use only Clifford algebras
(more properly speaking Clifford bundles), and do not use the concept of
\emph{extensor} are incomplete \cite{20}. These theories, e.g., cannot
capture the essential mathematical nature of the physical concepts of
energy-momentum and angular momentum associated with physical fields (in the
sense of Faraday), since they must be mathematically represented by extensor
fields. In particular, approaches to Dirac theory which do not use the
concept of extensors are incomplete, to say the less. On this issue, see (
\cite{14}-\cite{18}).} has been applied since a long time ago for
presentations of Maxwell theory, see e.g., (\cite{7}-\cite{10}), of Dirac
theory, see e.g., (\cite{11}-\cite{19}) and the theory of the gravitational
field, see e.g., (\cite{9},\cite{21},\cite{22}). Hestenes in 1966 \cite{7}
wrote a small book on the subject which has been source of inspiration for
many scientists and eventually spread unnecessary misconceptions on the
subject of Clifford algebras and their applications in physics. In 1984
Hestenes and Sobczyk published a book\footnote{%
This book is essentially based on Sobczyk Ph.D. thesis presented at the
Department of Mathematics of the University of Arizona. We are grateful to
Professor P. Lounesto (Helsinki) for this important information.},
emphasizing that Clifford algebras leads naturally to a \emph{geometric
calculus }\cite{23}. Some of the ideas that we will explore in papers I-VII
have their inspiration on that book. However, it must be emphasized that our
approach differs substantially from the one of those authors in many aspects
as the reader can verify. In particular, the theory of multivector
functionals and their derivatives with the full generality that the subject
deserves appears in this series for the first time. Our exposition of the
\emph{differential geometry} on manifolds (the subject of a new series of
papers), with a general theory of connections using the concept of extensor
fields is (we believe) a fresh approach to the subject. Our theory is \emph{%
not} based on the concept of vector manifolds used in \cite{23} (which
presents some problems \cite{24}) and can be applied rigorously to general
manifolds of arbitrary topology\footnote{%
A preliminary presentation of the general theory of connection using
multivector-extensor calculus on Minkowski manifolds appears in \cite{25}.}.
Our development of a Lagrangian theory of multivector fields \footnote{%
In the paper dealing with the Lagrangian formalism for fields we make use of
the concept of a spinor field that (roughly speaking) can be said to be an
equivalence class of a sum of non homogenous multivector fields. A tentative
definition of these objects appear in \cite{26}, which unfortunately
contains many misprints and some important errors. These are correct in (
\cite{31},\cite{32}).} improves preliminary presentations\footnote{%
See \cite{27} for a list of references on the subject.}, since now we give a
formulation valid for multivector fields and extensor fields over arbitrary
Lorentzian manifolds equipped with a general connection (\emph{not}
necessarily metric compatible)\footnote{%
It is also possible to present a theory of \emph{spinor} fields, where these
objects are (loosely speaking) represented by certain equivalence classes of
multivector fieds on an arbitrary manifold. A rigorous presentation of that
theory is given elsewhere (\cite{28},\cite{29}).}. We emphasize also that
our presentation of Einstein's gravitational theory (in a future series of
papers) using the multivector-extensor calculus on manifolds will
demonstrates that preliminary attempts [22] towards a theory of the
gravitational field based on these concepts is paved with some serious
mathematical (and also physical) misconceptions (see also [25] in this
respect) which invalidate them. We think that our presentation of the basic
working ideas about euclidean and metric Clifford algebras is reasonable
self complete for our purposes. However, there is still more concerning
Clifford algebra theory that has not been developed or even quoted in this
series of papers. These results are important for many applications ranging
from pure and applied mathematics to engineering and recent physical
theories (see, e.g., \cite{30} \cite{31}).

For readers that are newcomers to the subject we recommend the books by
Lounesto \cite{32} and Porteous (\cite{33},\cite{34}) for complementary
points of view and material relative to the developments that follows.

\section{The Euclidean Clifford Algebra of Multivectors}

Let $V$ be a vector space over $\mathbb{R}$ with finite
dimension, i.e., $\dim V=n,$ where $n\in N,$ and let $V^{*}$ be
the dual vector space to $V$. Recall that $\dim V=\dim V^{*}=n$.

Let $k$ be an integer number with $0\leq k\leq n$. The vector spaces of $k$%
-vectors and $k$-forms over $V$ as usual will be denoted by $\bigwedge^{k}V$
and $\bigwedge^{k}V^{*}$, respectively.\footnote{%
If the reader is not familiar with exterior algebra he must consult texts on
the subject. See, e.g., (\cite{35},\cite{36}\cite{37}). However, care must
be taken when reading different books which use \emph{different} definitions
for the \emph{exterior }product and still use all the same symbol for that
different products. About this issue, see comments on Appendix A.}

As well known, a $0$-vector can be identified with a real number, i.e., $%
\bigwedge^{0}V=\mathbb{R},$ an $1$-vector is the name of objects
living on $V,$ i.e., $\bigwedge^{1}V=V,$ and a $k$-vector with
$2\leq k\leq n$ is precisely a skew-symmetric contravariant
$k$-tensor over $V$. A $0$-form is also a real number, i.e.,
$\bigwedge^{0}V^{*}=\mathbb{R}$. A $1$-form is a form (or
covector) belonging to $V^{*},$ i.e., $\bigwedge^{1}V^{*}=V^{*},$ and a $k$%
-form with $2\leq k\leq n$ is exactly a skew-symmetric covariant $k$-tensor
over $V.$ Recall that $\dim \bigwedge^{k}V=\dim \bigwedge^{k}V^{*}=\binom{n}{%
k}$.

The $0$-vectors, $2$-vectors,$\ldots ,$ $(n-1)$-vectors and $n$-vectors are
sometimes called scalars, bivectors,$\ldots ,$ pseudovectors and
pseudoscalars, respectively. The $0$-forms, $2$-forms,$\ldots ,$ $(n-1)$%
-forms and $n$-forms are named as scalars, biforms,$\ldots ,$ pseudoforms
and pseudoscalars, respectively.

\subsection{Multivectors}

A formal sum of $k$-vectors over $V$ with $k$ running from $0$ to $n,$
\begin{equation}
X=X_{0}+X_{1}+\cdots +X_{n},  \label{1.1}
\end{equation}
is called a \emph{multivector} over $V.$

The set of multivectors over $V$ has a natural structure of
vector space over $\mathbb{R}$ and is usually denoted by
$\bigwedge V=\mathbb{R}+V+\cdots
+\bigwedge^{n}V.$ We have that $\dim \bigwedge V=\binom{n}{0}+\binom{n}{1}%
+\cdots +\binom{n}{n}=2^{n}.$

\subsubsection{$k$-Part, Grade Involution and Reversion Operators}

Let $k$ be an integer number with $0\leq k\leq n.$ The linear mapping $%
\bigwedge V\ni X\mapsto \left\langle X\right\rangle _{k}\in \bigwedge^{k}V$
such that for any $j$ with $0\leq j\leq n:$ if $j\neq k,$ then $\left\langle
X\right\rangle _{k}=0,$ i.e.,
\begin{equation}
\left\langle X\right\rangle _{k}=X_{k},  \label{1.2}
\end{equation}
for each $k=0,1,\cdots ,n,$ is called the $k$-part operator. $\left\langle
X\right\rangle _{k}$ is read as the $k$-part of $X.$

It is evident that any multivector can be written as sum of their $k$-parts
\begin{equation}
X=\sum_{k=0}^{n}\left\langle X\right\rangle _{k}. \label{1.3}
\end{equation}

There are several important automorphisms (or antiautomorphisms) on $%
\bigwedge V$. For what follows, we shall need to introduce some
automorphisms that are involutions on $\bigwedge V$. We have:\medskip

\textbf{i} The linear mapping $\bigwedge V\ni X\mapsto \hat{X}\in \bigwedge
V $ such that
\begin{equation}
\hat{X}=\sum_{k=0}^{n}(-1)^{k}\left\langle X\right\rangle _{k},
\label{1.4}
\end{equation}
is called the \emph{main automorphim} operator or \emph{grade involution}
operator. $\hat{X}$ is called the grade involution of $X.\medskip $

\textbf{ii} The linear mapping $\bigwedge V\ni X\mapsto \widetilde{X}\in
\bigwedge V$ such that
\begin{equation}
\widetilde{X}=\sum_{k=0}^{n}(-1)^{\frac{1}{2}%
k(k-1)}\left\langle X\right\rangle _{k},  \label{1.5}
\end{equation}
is an antiautomorphism called the \emph{reversion} operator. $\widetilde{X}$
is called the reverse of $X.$

Since the main automorphisms and reversion operators are involutions on the
vector space of multivectors, we have that $\widehat{\hat{X}}=X$ and $%
\widetilde{\widetilde{X}}=X.$ Both involutions commute with the $k$-part
operator, i.e., $\widehat{\langle X\rangle }_{k}=\left\langle \hat{X}%
\right\rangle _{k}$ and $\widetilde{\left\langle X\right\rangle _{k}}%
=\left\langle \widetilde{X}\right\rangle _{k},$ for each $k=0,1,\ldots ,n.$

The composition of the main automorphism with the reversion operator (in any
order) is called the \emph{conjugate} operator. The conjugate of $X$ will be
denoted by $\bar{X}$. We have
\begin{equation}
\bar{X}=\widetilde{\hat{X}}=\widehat{\tilde{X}}.  \label{1.5BIS}
\end{equation}

\subsection{Exterior Algebra}

We define the exterior product\footnote{%
There are several different definitions of the exterior product in the
literature differing by factors and all using the same symbol.. This may
lead to confusion if care is not taken. See Appendix A for some details.} of
$X_{p}\in \bigwedge^{p}V$ and $Y_{q}\in \bigwedge^{q}V$ by
\begin{equation}
X_{p}\wedge Y_{q}=\frac{(p+q)!}{p!q!}\mathcal{A}(X_{p}\otimes Y_{q}),
\label{1.6bis}
\end{equation}
where $X_{p}\otimes Y_{q}$ is the tensor product of $X_{p}$ by $Y_{q}$ (see
Appendix A) and $\mathcal{A}$ is the \emph{antisymmetrization operator},
i.e., a linear mapping $\mathcal{A}:T^{k}V\rightarrow \bigwedge^{k}V$ such
that

(\textbf{i}) for all $\alpha \in \mathbb{R}:\mathcal{A}\alpha
=\alpha ,$

(\textbf{ii}) for all $v\in V:\mathcal{A}v=v,$

(\textbf{iii}) for all $t\in T^{k}V,$ with $k\geq 2,$%
\begin{equation}
\mathcal{A}t(\omega ^{1},\ldots ,\omega ^{k})=\dfrac{1}{k!}\epsilon
_{i_{1}\ldots i_{k}}t(\omega ^{i_{1}},\ldots ,\omega ^{i_{k}}),
\label{1.6novo}
\end{equation}
where $\epsilon _{i_{1}\ldots i_{k}}$ is the permutation symbol of order $k,$

\begin{equation}
\epsilon _{i_{1}\ldots i_{k}}=\left\{
\begin{array}{cc}
1, & \text{if }i_{1}\ldots i_{k}\text{ is a even permutation of }1\ldots k
\\
-1, & \text{if }i_{1}\ldots i_{k}\text{ is odd permutation of }1\ldots k \\
0, & \text{otherwise}
\end{array}
\right.  \label{1.6novoo}
\end{equation}

Eq.(\ref{1.6bis}), with $p\geq 1$ and $q\geq 1$ means that for $\omega
^{1},\ldots ,\omega ^{p},\omega ^{p+1},\ldots ,\omega ^{p+q}\in V^{*},$%
\begin{eqnarray}
&&X_{p}\wedge Y_{q}(\omega ^{1},\ldots ,\omega ^{p},\omega ^{p+1},\ldots
,\omega ^{p+q})  \nonumber \\
&=&\frac{1}{p!q!}\epsilon _{i_{1}\ldots i_{p}i_{p+1}\ldots
i_{p+q}}X_{p}(\omega ^{i_{1}},\ldots ,\omega ^{i_{p}})Y_{q}(\omega
^{i_{p+1}},\ldots ,\omega ^{i_{p+q}}).  \label{1.6bisa}
\end{eqnarray}

From eq.(\ref{1.6bis}) by using a well-known property of the
antisymmetrization operator, namely: $\mathcal{A}(\mathcal{A}t\otimes u)=%
\mathcal{A}(t\otimes \mathcal{A}u)=\mathcal{A}(t\otimes u),$ a noticeable
formula for expressing simple $k$-vectors in terms of tensor products of $k$
vectors can be easily deduced. It is\footnote{%
Recall that $\epsilon ^{j_{1}\ldots j_{k}}\equiv \epsilon _{j_{1}\ldots
j_{k}}.$},
\begin{equation}
v_{1}\wedge \ldots \wedge v_{k}=\epsilon ^{j_{1}\ldots
j_{k}}v_{j_{1}}\otimes \ldots \otimes v_{j_{k}}.  \label{1.6biss}
\end{equation}
If $\omega ^{1},\ldots ,\omega ^{k}\in V^{*},$ then
\begin{equation}
v_{1}\wedge \ldots \wedge v_{k}(\omega ^{1},\ldots ,\omega ^{k})=\epsilon
^{j_{1}\ldots j_{k}}\omega ^{1}(v_{j_{1}})\ldots \omega ^{k}(v_{j_{k}}).
\label{1.6bisss}
\end{equation}

Now, we define the exterior product of \emph{multivectors} $X$ and $Y$ as
being the mutivector with components $\left\langle X\wedge Y\right\rangle
_{k}$ such that
\begin{equation}
\left\langle X\wedge Y\right\rangle _{k}=\sum_{j=0}^{k}%
\left\langle X\right\rangle _{j}\wedge \left\langle
Y\right\rangle _{k-j}, \label{1.6}
\end{equation}
for each $k=0,1,\ldots ,n.$ Note that on the right side there appears the
exterior product of $j$-vectors and $(k-j)$-vectors with $0\leq j\leq n$.

This exterior product is an \emph{internal} composition law on $\bigwedge V$%
. It is associative and satisfies the usual distributive laws (on the left
and on the right).

The vector space $\bigwedge V$ endowed with this exterior product $\wedge $
is an associative algebra called the \emph{exterior algebra} of multivectors.

We recall now for future use some important properties of the exterior
algebra of multivectors:

\textbf{ei} For any $\alpha ,\beta \in \mathbb{R},$ $X\in \bigwedge V$%
\begin{eqnarray}
\alpha \wedge \beta &=&\beta \wedge \alpha =\alpha \beta \text{ (real
product),}  \label{1.7a} \\
\alpha \wedge X &=&X\wedge \alpha =\alpha X\text{ (multiplication by
scalars).}  \label{1.7b}
\end{eqnarray}

\textbf{eii} For any $X_{j}\in \bigwedge^{j}V$ and $Y_{k}\in \bigwedge^{k}V$%
\begin{equation}
X_{j}\wedge Y_{k}=(-1)^{jk}Y_{k}\wedge X_{j}.  \label{1.8}
\end{equation}

\textbf{eiii} For any $X,Y\in \bigwedge V$%
\begin{eqnarray}
\widehat{X\wedge Y} &=&\hat{X}\wedge \hat{Y},  \label{1.9a} \\
\widetilde{X\wedge Y} &=&\widetilde{Y}\wedge \widetilde{X}.  \label{1.9b}
\end{eqnarray}

\subsection{Euclidean Scalar Product}

Let $G_{E}$ be an \emph{euclidean metric} for $V$, i.e., a mapping $%
G_{E}:V\times V\rightarrow \mathbb{R}$ which is a symmetric,
non-degenerate and positive definite bilinear form over $V,$
\begin{eqnarray}
G_{E}(v,w) &=&G_{E}(w,v)\text{ }\forall v,w\in V  \label{1.EUCL1} \\
\text{If }G_{E}(v,w) &=&0\text{ }\forall w\in V,\text{ then }v=0
\label{1.EUCL2} \\
G_{E}(v,v) &\geqslant &0\text{ }\forall v\in V\text{ and if }G_{E}(v,v)=0,%
\text{ then }v=0.  \label{1.EUCL3}
\end{eqnarray}

It is usual to write
\begin{equation}
G_{E}(v,w)\equiv v\cdot w  \label{1.EUCL1BIS}
\end{equation}
where $v\cdot w$ is said to be the \emph{scalar product} of the vectors $%
v,w\in V$.

This practice forgets that any scalar product is relative to a
given $G_{E},$ it is a fact which will be important for the
developments that follows, the correct notation should be
$v\underset{G_{E}}{\cdot }w$. Nevertheless, when no confusion
arises we will follow the standard practice.

The pair $(V,G_{E})$ is called an \emph{euclidean structure} for $V$.
Sometimes, an euclidean structure is also called an \emph{euclidean space}.
It is very important to realize that there are an infinite of euclidean
structures for a real vector space $V$. Two euclidean structures $(V,G_{E})$
and $(V,G_{E}^{\prime })$ are equal if and only if $G_{E}=G_{E}^{\prime }$.

Let $\mathfrak{B}$ be the set of all basis of $V$. It means that a
generic element of $\mathfrak{B}$ is an \emph{ordered} set of
linearly independent vectors of $V$, say
$(e_{1},e_{2},...,e_{n}),$ which will be denoted simply by
$\{e_{k}\}$ in what follows.

Now, given an euclidean structure for $V$, we can immediately
select a subset $\mathfrak{B}_{O}$ of $\mathfrak{B}$ whose
elements are of the orthonormal bases according to the euclidean
structure. This means that if $\{f_{k}\}\in \mathfrak{B}_{O}$,
then
\begin{equation}
G_{E}(f_{i},f_{j})\equiv f_{i}\cdot f_{j}=\delta _{ij},  \label{1.EUCL4}
\end{equation}
where $\delta _{ij}=\left\{
\begin{array}{cc}
1, & i=j=1,2,...,n \\
0, & i\neq j
\end{array}
\right. .$ It is trivial to realize that any two basis $\{f_{k}\},\{f_{k}^{%
\prime }\}\in \mathfrak{B}_{O}$ are related by a linear orthogonal
transformation, i.e., $f_{k}^{\prime }=\left. L_{k}\right.
^{i}f_{i},$ where the matrix $\mathbf{L}$\textbf{\ }whose entries
are the real numbers $\left.
L_{k}\right. ^{i}$ is orthogonal, i.e., $\mathbf{L}^{t}\mathbf{L=\mathbf{LL}%
^{t}=1}.$

Once an euclidean structure $(V,G_{E})$ has been set we can equip $%
\bigwedge^{p}V$ with an \emph{euclidean scalar product of }$p$-\emph{vectors}%
. $\bigwedge V$ can be endowed with an \emph{euclidean} \emph{scalar product
of multivectors.}

Let $\{e_{k}\}$ be any basis of $V,$ and $\{\varepsilon ^{k}\}$
be the dual basis of $\{e_{k}\}.$ As we know, $\{\varepsilon
^{k}\}$ is the unique basis of $V^{*}$ such that $\varepsilon
^{k}(e_{j})=\delta _{j}^{k}.$ Associated to $(V,G_{E})$ we define
the scalar product of $p$-vectors $X_{p},Y_{p}\in
\bigwedge^{p}V,$ namely $X_{p}\cdot Y_{p}\in \mathbb{R},$ by the
following axioms:

\textbf{Ax-i} For all $\alpha ,\beta \in \mathbb{R},$
\begin{equation}
\alpha \cdot \beta =\alpha \beta \text{ (real product).}  \label{1.EUCL5a}
\end{equation}

\textbf{Ax-ii} For all $X_{p},Y_{p}\in \bigwedge^{p}V$ with $p\geq 1,$
\begin{equation}
X_{p}\cdot Y_{p}=(\frac{1}{p!})^{2}X_{p}(\varepsilon ^{I})Y_{p}(\varepsilon
^{J})\det \left[ G_{E}(e_{I},e_{J})\right] ,  \label{1.EUCL5b}
\end{equation}

where we use (conveniently) the short notations
\begin{eqnarray}
X_{p}(\varepsilon ^{I}) &\equiv &X_{p}(\varepsilon ^{i_{1}},\ldots
,\varepsilon ^{i_{p}})=\underset{p}{X}^{i_{1}...i_{p}},
\label{1.EUCL6a}
\\
Y_{p}(\varepsilon ^{J}) &\equiv &Y_{p}(\varepsilon ^{j_{1}},\ldots
,\varepsilon ^{j_{p}})=\underset{p}{Y}^{j_{1}...j_{p}}.
\label{1.EUCL6b}
\end{eqnarray}
$X_{p}(\varepsilon ^{i_{1}},\ldots ,\varepsilon ^{i_{p}})$ and $%
Y_{p}(\varepsilon ^{j_{1}},\ldots ,\varepsilon ^{j_{p}})$ are the components
of $X_{p}$ and $Y_{p}$ with respect to the $p$-vector basis $%
\{e_{j_{1}}\wedge \ldots \wedge e_{j_{p}}\}$ and $1\leq j_{1}<\cdots
j_{p}\leq n,$ i.e.,
\begin{equation}
X_{p}=\frac{1}{p!}X_{p}(\varepsilon ^{i_{1}},\ldots ,\varepsilon
^{i_{p}})e_{i_{1}}\wedge \ldots e_{i_{p}}\text{ and }Y_{p}=\frac{1}{p!}%
Y_{p}(\varepsilon ^{j_{1}},\ldots ,\varepsilon ^{j_{p}})e_{j_{1}}\wedge
\ldots e_{j_{p}}.  \label{1.EUCL6c}
\end{equation}
Also,
\begin{equation}
\det \left[ G_{E}(e_{I},e_{J})\right] \equiv \det \left[
\begin{array}{ccc}
G_{E}(e_{i_{1}},e_{j_{1}}) & \ldots & G_{E}(e_{i_{1}},e_{j_{k}}) \\
\ldots & \ldots & \ldots \\
G_{E}(e_{i_{k}},e_{j_{1}}) & \ldots & G_{E}(e_{i_{k}},e_{j_{k}})
\end{array}
\right] .  \label{1.EUCL6d}
\end{equation}
Note that in eq.(\ref{1.EUCL5b}) the Einstein \emph{convention} for sums
over the indices $I\equiv i_{1},\ldots ,i_{p}=1,\ldots ,n$ and $J\equiv
j_{1},\ldots ,j_{p}=1,\ldots ,n$ was used.

It is not difficult to realize that the scalar product defined by the axioms
\textbf{i}-\textbf{ii} does not depend on the bases $\{e_{k}\}$ and $%
\{\varepsilon ^{k}\}$ for calculating it.

It is a well-defined \emph{euclidean scalar product on} $\bigwedge^{p}V,$
since it is symmetric, satisfies the distributive laws, has the mixed
associative property and is non-degenerate, i.e., if $X_{p}\cdot Y_{p}=0$
fot all $Y_{p},$ then $X_{p}=0.$ It is also satisfying the strong property
of being \emph{positive definite}, i.e., $X_{p}\cdot X_{p}\geq 0$ for all $%
X_{p}$ and if $X_{p}\cdot X_{p}=0,$ then $X_{p}=0.$

So the scalar product on $\bigwedge^{p}V$ as defined by eqs.(\ref{1.EUCL5a})
and (\ref{1.EUCL5b}) will be called the euclidean scalar product of $p$%
-vectors associated to $(V,G_{E}).$

Now, associated to $(V,G_{E})$ we define the scalar product of multivectors $%
X,Y\in \bigwedge V,$ namely $X\cdot Y\in \mathbb{R},$ by
\begin{equation}
X\cdot Y=\sum_ {k=0}^{n}\left\langle X\right\rangle _{k}\cdot
\left\langle Y\right\rangle _{k}. \label{1.2.EUCL0}
\end{equation}
Note that on the right side there appears the scalar product of $k$-vectors
with $0\leq k\leq n,$ as defined by eqs.(\ref{1.EUCL5a}) and (\ref{1.EUCL5b}%
).

By using eqs.(\ref{1.EUCL5a}) and (\ref{1.EUCL5b}) we can easily note that
eq.(\ref{1.2.EUCL0}) can be written as

\begin{equation}
X\cdot Y=\left\langle X\right\rangle _{0}\left\langle Y\right\rangle _{0}+%
\sum_{k=1}^{n} (\frac{1}{k!})^{2}\left\langle X\right\rangle
_{k}(\varepsilon ^{I})\left\langle Y\right\rangle
_{k}(\varepsilon ^{J})\det \left[ G_{E}(e_{I},e_{J})\right] .
\label{1.2.EUCL1}
\end{equation}
Recall that in eq.(\ref{1.2.EUCL1}) the Einstein \emph{convention} for sums
over the indices $I\equiv i_{1},\ldots ,i_{k}=1,\ldots ,n$ and $J\equiv
j_{1},\ldots ,j_{k}=1,\ldots ,n$ was used.

It is very important here to notice that the scalar product as defined by
eq.(\ref{1.2.EUCL0}) is a well-defined \emph{euclidean scalar product on }$%
\bigwedge V.$ It is symmetric, satisfies the distributive laws, has the
mixed associative property and is non-degenerate, i.e., if $X\cdot Y=0$ for
all $Y,$ then $X=0.$ In addition, it has also the strong property of being
\emph{positive definite,} i.e., $X\cdot X\geq 0$ for all $X$ and if $X\cdot
X=0,$ then $X=0.$

So the scalar product on $\bigwedge V$ as defined by eq.(\ref{1.2.EUCL0})
will be called the euclidean scalar product of multivectors associated to $%
(V,G_{E}).$

Now, note that if we take any orthonormal basis $\{f_{k}\}$ with respect to $%
(V,G_{E}),$ i.e., $f_{j}\cdot f_{k}=\delta _{jk},$ whose dual basis is $%
\{\varphi ^{k}\},$ i.e., $\varphi ^{k}(f_{j})=\delta _{j}^{k},$ we have that
$\det [G_{E}(f_{I},,f_{J})]=\epsilon _{i_{1}\ldots i_{k}}^{j_{1}\ldots
j_{k}}=\epsilon _{j_{1}\ldots j_{k}}^{i_{1}\ldots i_{k}}.$ Then, by taking
into account\footnote{%
Recall that $\epsilon _{i_{1}\ldots i_{k}}^{j_{1}\ldots j_{k}}$ is the
so-called generalized permutation symbol of order $k,$
\[
\epsilon _{i_{1}\ldots i_{k}}^{j_{1}\ldots j_{k}}=\det \left[
\begin{array}{ccc}
\delta _{i_{1}}^{j_{1}} & \ldots & \delta _{i_{1}}^{j_{k}} \\
\ldots & \ldots & \ldots \\
\delta _{i_{k}}^{j_{1}} & \ldots & \delta _{i_{k}}^{j_{k}}
\end{array}
\right] ,\text{ with }i_{1},\ldots ,i_{k}=1,\ldots ,n\text{ and }%
j_{1},\ldots ,j_{k}=1,\ldots ,n.
\]
} that $\dfrac{1}{k!}\epsilon _{i_{1}\ldots i_{k}}^{j_{1}\ldots
j_{k}}\left\langle X\right\rangle _{k}(\varphi ^{i_{1}},\ldots ,\varphi
^{i_{k}})=\left\langle X\right\rangle _{k}(\varphi ^{j_{1}},\ldots ,\varphi
^{j_{k}}),$ we can easily see that eq.(\ref{1.2.EUCL1}) can be written as
\begin{equation}
X\cdot Y=\left\langle X\right\rangle _{0}\left\langle Y\right\rangle _{0}+%
\sum_{k=1}^{n}\frac{1}{k!}\sum_{j_{1}\ldots j_{k}=1}%
{n}\left\langle X\right\rangle _{k}(\varphi ^{j_{1}},\ldots
,\varphi ^{j_{k}})\left\langle Y\right\rangle _{k}(\varphi
^{j_{1}},\ldots ,\varphi ^{j_{k}}). \label{1.2.EUCL2}
\end{equation}

It should be noted that eq.(\ref{1.2.EUCL2}) in the particular case of
vectors is reduced to
\begin{equation}
v\cdot w=\sum_{j=1}^{n}\varphi ^{j}(v)\varphi ^{j}(w).
\label{1.2.EUCL2a}
\end{equation}

We summarize now the basic properties of the euclidean scalar product of
multivectors.

\textbf{esi} For any $\alpha ,\beta \in \mathbb{R}:$%
\begin{equation}
\alpha \cdot \beta =\alpha \beta \text{ (real product).}  \label{1.0.1}
\end{equation}

\textbf{esii} For any $v,w\in V:$%
\begin{equation}
v\cdot w=G_{E}(v,w).  \label{1.0.2}
\end{equation}
It shows that eq.(\ref{1.2.EUCL0}) contains the scalar product of vectors.

\textbf{esiii} For any $X_{j}\in \bigwedge^{j}V$ and $Y_{k}\in
\bigwedge^{k}V:$%
\begin{equation}
X_{j}\cdot Y_{k}=0,\text{ if }j\neq k.  \label{1.0.3}
\end{equation}

The properties given by eq.(\ref{1.0.1}), eq.(\ref{1.0.2}) and eq.(\ref
{1.0.3}) follow directly from the definition given by eq.(\ref{1.2.EUCL0}).

\textbf{esiv} For any simple $k$-vectors $v_{1}\wedge \ldots v_{k}\in
\bigwedge^{k}V$ and $w_{1}\wedge \ldots w_{k}\in \bigwedge^{k}V:$%
\begin{equation}
(v_{1}\wedge \ldots v_{k})\cdot (w_{1}\wedge \ldots w_{k})=\det \left[
\begin{array}{ccc}
v_{1}\cdot w_{1} & \ldots & v_{1}\cdot w_{k} \\
\ldots & \ldots & \ldots \\
v_{k}\cdot w_{1} & \ldots & v_{k}\cdot w_{k}
\end{array}
\right] .  \label{1.0.4}
\end{equation}
\begin{proof}
We will use eq.(\ref{1.2.EUCL2}). Then, by using eq.(\ref{1.6bisss}) and eq.(%
\ref{1.2.EUCL2a}), and recalling the $k\times k$ determinant formula, $\det
\left[ a_{pq}\right] =\dfrac{1}{k!}\epsilon ^{p_{1}\ldots p_{k}}\epsilon
^{q_{1}\ldots q_{k}}a_{p_{1}q_{1}}\ldots a_{p_{k}q_{k}},$ we have
\begin{eqnarray*}
&&(v_{1}\wedge \ldots v_{k})\cdot (w_{1}\wedge \ldots w_{k}) \\
&=&\frac{1}{k!}\sum_{j_{1}\ldots j_{k}=1 }^{n}%
(v_{1}\wedge \ldots v_{k})(\varphi ^{j_{1}},\ldots ,\varphi
^{j_{k}})(w_{1}\wedge \ldots w_{k})(\varphi ^{j_{1}},\ldots ,\varphi
^{j_{k}}) \\
&=&\frac{1}{k!} \sum_{j_{1}\ldots j_{k}=1}^{n} \epsilon
^{p_{1}\ldots p_{k}}\epsilon ^{q_{1}\ldots q_{k}}\varphi
^{j_{1}}(v_{p_{1}})\ldots \varphi ^{j_{k}}(v_{p_{k}})\varphi
^{j_{1}}(w_{q_{1}})\ldots \varphi ^{j_{k}}(w_{q_{k}}) \\
&=&\frac{1}{k!}\epsilon ^{p_{1}\ldots p_{k}}\epsilon ^{q_{1}\ldots q_{k}}%
\sum_{j_{1}=1}^{n}\varphi ^{j_{1}}(v_{p_{1}})\varphi
^{j_{1}}(w_{q_{1}})\ldots \sum_{j_{k}=1}^ {n}\varphi
^{j_{k}}(v_{p_{k}})\varphi ^{j_{k}}(w_{q_{k}}) \\
&=&\frac{1}{k!}\epsilon ^{p_{1}\ldots p_{k}}\epsilon ^{q_{1}\ldots
q_{k}}(v_{p_{1}}\cdot w_{q_{1}})\ldots (v_{p_{k}}\cdot w_{q_{k}}), \\
&=&\det \left[ v_{p}\cdot w_{q}\right] .
\end{eqnarray*}
\end{proof}

\begin{proposition}
Let $(\{e_{k}\},\{e^{k}\})$ be any pair of euclidean reciprocal bases of $V,$
i.e., $e_{k}\cdot e^{l}\equiv G_{E}(e_{k},e^{l})=\delta _{k}^{l}.$ For all $%
X\in \bigwedge V$ we have the following two expansion formulas
\begin{eqnarray}
X &=&X\cdot 1+\sum_{k=1}^{n}\frac{1}{k!}X\cdot (e^{j_{1}}\wedge
\ldots e^{j_{k}})(e_{j_{1}}\wedge \ldots e_{j_{k}})
\label{1.13a} \\
X &=&X\cdot 1+\sum_{k=1}^{n}\frac{1}{k!}X\cdot (e_{j_{1}}\wedge
\ldots e_{j_{k}})(e^{j_{1}}\wedge \ldots e^{j_{k}}). \label{1.13b}
\end{eqnarray}
\end{proposition}

\begin{proof}
We give here the proof for vectors and $p$-vectors, with $p\geq
2$.
For $v\in V,$ since $\{e_{k}\}$ and $\{e^{k}\}$ are bases of
$V,$ there are unique real numbers $v^{i}$ and $v_{i}$ with
$i=1,\ldots ,n$ such that
\[
v=v^{i}e_{i}=v_{i}e^{i}.
\]

Let us calculate $v\cdot e^{j}$ and $v\cdot e_{j}.$ Then, by taking into
account the reciprocity condition of $(\{e_{k}\},\{e^{k}\})$, we get
\[
v=(v\cdot e^{j})e_{j}=(v\cdot e_{j})e^{j}.
\]
It is standard practice to call $v\cdot e^{j}$ and $v\cdot e_{j}$
respectively the contravariant and covariant $j$-th components of $v$.

For $X\in \bigwedge^{p}V,$ there are unique real numbers $X^{i_{1}\ldots
i_{p}}$ and $X_{i_{1}\ldots i_{p}}$ with $i_{1},\ldots ,i_{p}=1,\ldots ,n$
such that
\begin{equation}
X=\frac{1}{p!}X^{i_{1}\ldots i_{p}}e_{i_{1}}\wedge \ldots e_{i_{p}}=\frac{1}{%
p!}X_{i_{1}\ldots i_{p}}e^{i_{1}}\wedge \ldots e^{i_{p}}.  \label{1.13c}
\end{equation}

Then, by taking for example the scalar products $X\cdot (e^{j_{1}}\wedge
\ldots e^{j_{p}})$. By using eq.(\ref{1.0.4}), the reciprocity condition of $%
(\{e_{k}\},\{e^{k}\})$ and the combinatorial formula $\dfrac{1}{p!}\epsilon
_{i_{1}\ldots i_{p}}^{j_{1}\ldots j_{p}}X^{i_{1}\ldots i_{p}}=X^{j_{1}\ldots
j_{p}},$ we have
\begin{eqnarray*}
X\cdot (e^{j_{1}}\wedge \ldots e^{j_{p}}) &=&\frac{1}{p!}X^{i_{1}\ldots
i_{p}}(e_{i_{1}}\wedge \ldots e_{i_{p}})\cdot (e^{j_{1}}\wedge \ldots
e^{j_{p}}) \\
&=&\frac{1}{p!}X^{i_{1}\ldots i_{p}}\det \left[
\begin{array}{ccc}
e_{i_{1}}\cdot e^{j_{1}} & \ldots & e_{i_{1}}\cdot e^{j_{p}} \\
\ldots & \ldots & \ldots \\
e_{i_{p}}\cdot e^{j_{1}} & \ldots & e_{i_{p}}\cdot e^{j_{p}}
\end{array}
\right] \\
&=&\frac{1}{p!}\epsilon _{i_{1}\ldots i_{p}}^{j_{1}\ldots
j_{p}}X^{i_{1}\ldots i_{p}}=X^{j_{1}\ldots j_{p}},
\end{eqnarray*}
i.e., $X^{j_{1}\ldots j_{p}}=X\cdot (e^{j_{1}}\wedge \ldots e^{j_{p}}).$
Analogously, we can prove that $X_{j_{1}\ldots j_{p}}=X\cdot
(e_{j_{1}}\wedge \ldots e_{j_{p}})$.

Then, we get
\[
X=\frac{1}{p!}X\cdot (e^{j_{1}}\wedge \ldots e^{j_{p}})e_{j_{1}}\wedge
\ldots e_{j_{p}}=\frac{1}{p!}X\cdot (e_{j_{1}}\wedge \ldots
e_{j_{p}})e^{j_{1}}\wedge \ldots e^{j_{p}}.
\]

Hence, eqs.(\ref{1.13a}) and (\ref{1.13b}) follows from the statement above
and essentially from eq.(\ref{1.0.3}).%

\end{proof}

\subsection{$b$-Metric}

Let $\{b_{k}\}$ be any but fixed basis of $V,$ and let $\{\beta
^{k}\}$ be a basis of $V^{*}$dual to $\{b_{k}\},$ i.e., $\beta
^{k}(b_{j})=\delta _{j}^{k}.$ Associated to $\{b_{k}\}$ we can
introduce an euclidean metric on $V,$ say $\underset{b}{G_{E}},$
defined by
\begin{equation}
\underset{b}{G_{E}}(v,w)=\delta _{jk}\beta ^{j}(v)\beta ^{k}(w),
\label{1.1a}
\end{equation}
i.e., $\underset{b}{G_{E}}=\delta _{jk}\beta ^{j}\otimes \beta
^{k}.$

It is a well defined euclidean metric on $V,$ since $\underset{b}{G_{E}}%
\in T_{2}(V)$ is symmetric non-degenerate and positive definite,
as it is easy to verify. Such $\underset{b}{G_{E}}$ will be
called a \emph{fiducial
metric on }$V$\emph{\ induced by }$\{b_{k}\},$ or for short, a $b$\emph{%
-metric. }The euclidean structure $(V,\underset{b}{G_{E}})$ will
be called a \emph{fiducial metric structure for }$V$\emph{\
induced by }$\{b_{k}\},$ or for short, a $b$\emph{-metric
structure. }The pair $(V,\{b_{k}\})$ could
be called a \emph{fiducial structure for }$V$\emph{\ associated to }$%
\{b_{k}\},$ or for short, a $b$\emph{-structure.}

On another way of thinking we are equipping $V$ with a positive definite
scalar product of vectors naturally induced by $\{b_{k}\}.$ We write
\begin{equation}
v\underset{b}{\cdot }w\equiv \underset{b}{G_{E}}(v,w).
\label{1.1b}
\end{equation}

We present now two remarkable properties of a $b$-metric structure.

\textbf{i} The basis $\{b_{k}\}$ is orthonormal with respect to $(V,%
\underset{b}{G_{E}}),$ i.e.,
\begin{equation}
b_{j}\underset{b}{\cdot }b_{k}=\delta _{jk}. \label{1.1c}
\end{equation}

\textbf{ii} The scalar product of multivectors associated to $(V,\underset{%
b}{G_{E}})$ is given by the noticeable formula
\begin{equation}
X\underset{b}{\cdot }Y=\left\langle X\right\rangle
_{0}\left\langle
Y\right\rangle _{0}+\sum_ {k=1}^{n}\frac{1}{k!}%
\sum_{j_{1}\ldots j_{k}=1}^{n} \left\langle X\right\rangle
_{k}(\beta ^{j_{1}},\ldots ,\beta ^{j_{k}})\left\langle
Y\right\rangle _{k}(\beta ^{j_{1}},\ldots ,\beta ^{j_{k}}).
\label{1.1d}
\end{equation}

We know that all $b$-metric structure is a well-defined euclidean
structure. However, it might as well be asked if any euclidean
structure $(V,G_{E})$ is some $b$-metric structure
$(V,\underset{b}{G_{E}}).$ The answer is YES.

Given an euclidean metric $G_{E},$ by the Gram-Schmidt procedure, there is
an orthonormal basis $\{b_{k}\}$ with respect to $(V,G_{E}),$ i.e., $%
b_{j}\cdot b_{k}\equiv G_{E}(b_{j},b_{k})=\delta _{jk},$ such that the $b$%
-metric $\underset{b}{G_{E}}$ induced by $\{b_{k}\}$ coincides with $%
G_{E}. $ Indeed, if $\{\beta ^{k}\}$ is the dual basis of $\{b_{k}\},$ then
\begin{eqnarray*}
\underset{b}{G_{E}}(v,w) &=&\delta _{jk}\beta ^{j}(v)\beta
^{k}(w)=G_{E}(b_{j},b_{k})\beta ^{j}(v)\beta ^{k}(w) \\
&=&G_{E}(\beta ^{j}(v)b_{j},\beta ^{k}(w)b_{k})=G_{E}(v,w),
\end{eqnarray*}
i.e., $\underset{b}{G_{E}}=G_{E}$

\subsubsection{$b$-Reciprocal Bases}

Let $\{e_{k}\}$ be any basis of $V,$ and $\{\varepsilon ^{k}\}$ be its dual
basis of $V^{*},$ i.e., $\varepsilon ^{k}(e_{j})=\delta _{j}^{k}.$ Let us
take a $b$-metric structure $(V,\underset{b}{G_{E}}).$ Associated to $%
\{e_{k}\},$ it is possible to define another basis for $V,$ say $\{e^{k}\},$
given by
\begin{equation}
e^{k}=\sum_{j=1}^{n}\varepsilon ^{k}(b_{j})b_{j}. \label{1.13.1}
\end{equation}
Since the set of the $n$ forms $\varepsilon ^{1},\ldots ,\varepsilon ^{n}\in
V^{*},$ is a basis for $V^{*}$, they are linearly independent covectors. It
follows that the $n$ vectors $e^{1},\ldots ,e^{n}\in V$ are also linearly
independent and constitutes a well-defined basis for $V.$

\begin{proposition}
The bases $\{e_{k}\}$ and $\{e^{k}\}$ satisfy the following $b$-scalar
product conditions
\begin{equation}
e_{k}\underset{b}{\cdot }e^{l}=\delta _{k}^{l}. \label{1.13.2}
\end{equation}
\end{proposition}

\begin{proof}
Using eqs.(\ref{1.13.1}) and (\ref{1.1c}), and the duality condition of $%
(\{e_{k}\},\{\varepsilon ^{k}\})$ we have
\[
e_{k}\underset{b}{\cdot }e^{l}=\sum_ {j=1}^{n}\varepsilon ^{l}(b_{j})(e_{k}\underset{b}{\cdot }b_{j})=\varepsilon ^{l}(%
\sum_{j=1}^{n}(e_{k}\underset{b}{\cdot }%
b_{j})b_{j})=\varepsilon ^{l}(e_{k})=\delta _{k}^{l}.%
\]
\end{proof}

It is noticeable that $\{e^{k}\}$ given by eq.(\ref{1.13.1}) is the unique
basis of $V$ which satisfies eq.(\ref{1.13.2}). Such a basis $\{e^{k}\}$
will be called the $\emph{b}$-\emph{reciprocal basis} of $\{e_{k}\}$. In
what follows we say that $\{e_{k}\}$ and $\{e^{k}\}$ are $b$-reciprocal
bases to each other.

In particular, the $b$-reciprocal basis of $\{b_{k}\}$ is itself, i.e.,
\begin{equation}
b^{k}=b_{k}\text{ for each }k=1,\ldots ,n,  \label{1.13.3}
\end{equation}
It follows directly from eq.(\ref{1.13.1}) and the duality condition of $%
(\{b_{k}\},\{\beta ^{k}\})$.

\subsection{Euclidean Interior Algebras}

Let us take an euclidean structure $(V,G_{E}).$ We can define two kind of
\emph{contracted products} for multivectors, namely $\lrcorner $ and $%
\llcorner $. If $X,Y\in \bigwedge V$ then $X\lrcorner Y\in \bigwedge V$ and $%
X\llcorner Y\in \bigwedge V$ such that
\begin{eqnarray}
(X\lrcorner Y)\cdot Z &=&Y\cdot (\widetilde{X}\wedge Z)  \label{1.17a} \\
(X\llcorner Y)\cdot Z &=&X\cdot (Z\wedge \widetilde{Y}),  \label{1.17b}
\end{eqnarray}
for all $Z\in \bigwedge V$.

These contracted products $\lrcorner $ and $\llcorner $ are internal laws on
$\bigwedge V.$ Both contracted products satisfy distributive laws (on the
left and on the right) but they are \emph{not} associative.

The vector space $\bigwedge V$ endowed with each of these contracted
products (either $\lrcorner $ or $\llcorner $) is a non-associative algebra.
They are called the \emph{euclidean interior algebras of multivectors.}

We present now some of the most important properties of the contracted
products.\medskip

\textbf{eip-i} For any $\alpha ,\beta \in \mathbb{R}$ and $X\in \bigwedge V:$%
\begin{eqnarray}
\alpha \lrcorner \beta &=&\alpha \llcorner \beta =\alpha \beta \text{ (real
product),}  \label{1.18a} \\
\alpha \lrcorner X &=&X\llcorner \alpha =\alpha X\text{ (multiplication by
scalars).}  \label{1.18b}
\end{eqnarray}

\textbf{eip-ii} For any $X_{j}\in \bigwedge^{j}V$ and $Y_{k}\in
\bigwedge^{k}V$ $(j\leq k):$%
\begin{equation}
X_{j}\lrcorner Y_{k}=(-1)^{j(k-j)}Y_{k}\llcorner X_{j}\text{.}  \label{1.19}
\end{equation}

\textbf{eip-iii} For any $X_{j}\in \bigwedge^{j}V$ and $Y_{k}\in
\bigwedge^{k}V:$%
\begin{eqnarray}
X_{j}\lrcorner Y_{k} &=&0,\text{ if }j>k,  \label{1.20a} \\
X_{j}\llcorner Y_{k} &=&0,\text{ if }j<k.  \label{1.20b}
\end{eqnarray}

\textbf{eip-iv} For any $X_{k},Y_{k}\in \bigwedge^{k}V:$%
\begin{equation}
X_{k}\lrcorner Y_{k}=X_{k}\llcorner Y_{k}=\widetilde{X_{k}}\cdot
Y_{k}=X_{k}\cdot \widetilde{Y_{k}}.  \label{1.21}
\end{equation}

\textbf{eip-v} For any $v\in V$ and $X,Y\in \bigwedge V:$%
\begin{equation}
v\lrcorner (X\wedge Y)=(v\lrcorner X)\wedge Y+\widehat{X}\wedge (v\lrcorner
Y).  \label{1.22}
\end{equation}

\subsection{Euclidean Clifford Algebra}

We define now an \emph{euclidean} \emph{Clifford product} of multivectors $X$
and $Y$ relative to a given euclidean structure $(V,G_{E}),$ denoted by
juxtaposition, by the following axioms:\medskip

\textbf{Ax-i} For all $\alpha \in \mathbb{R}$ and $X\in \bigwedge
V:\alpha X=X\alpha $ equals the multiplication of multivector $X$
by scalar $\alpha .\medskip $

\textbf{Ax-ii} For all $v\in V$ and $X\in \bigwedge V:vX=v\lrcorner
X+v\wedge X$ and $Xv=X\llcorner v+X\wedge v.\medskip $

\textbf{Ax-iii} For all $X,Y,Z\in \bigwedge V:X(YZ)=(XY)Z.\medskip $

The Clifford product is an internal law on $\bigwedge V.$ It is \emph{%
associative} (by the axiom (Ax-iii)) and satisfies distributive laws (on the
left and on the right). The distributive laws follow from the corresponding
distributive laws of the contracted and exterior products.

The vector space of multivectors over $V$ endowed with the Clifford product
is an associative algebra. It will be called \emph{euclidean Clifford
algebra of multivectors }and denoted by $\mathcal{C\ell }(V,G_{E})$%
.

Some important formulas which hold in $\mathcal{C\ell }(V,G_{E})$
are the following.

\textbf{eca-i }For any $v\in V$ and $X\in \bigwedge V:$%
\begin{eqnarray}
v\lrcorner X &=&\frac{1}{2}(vX-\widehat{X}v)  \label{1.23a} \\
\text{and }X\llcorner v &=&\frac{1}{2}(Xv-v\widehat{X}).  \nonumber \\
v\wedge X &=&\frac{1}{2}(vX+\widehat{X}v)  \label{1.23b} \\
\text{and }X\wedge v &=&\frac{1}{2}(Xv+v\widehat{X}).  \nonumber
\end{eqnarray}

\textbf{eca-ii }For any $X,Y\in \bigwedge V:$%
\begin{equation}
X\cdot Y=\left\langle \widetilde{X}Y\right\rangle _{0}=\left\langle X%
\widetilde{Y}\right\rangle _{0}.  \label{1.24}
\end{equation}

\textbf{eca-iii }For any $X,Y,Z\in \bigwedge V:$%
\begin{eqnarray}
(XY)\cdot Z &=&Y\cdot (\widetilde{X}Z)=X\cdot (Z\widetilde{Y}),
\label{1.25a} \\
X\cdot (YZ) &=&(\widetilde{Y}X)\cdot Z=(X\widetilde{Z})\cdot Y.
\label{1.25b}
\end{eqnarray}

\textbf{eca-iv }For any $X,Y\in \bigwedge V:$%
\begin{eqnarray}
\widehat{XY} &=&\widehat{X}\text{ }\widehat{Y},  \label{1.25c} \\
\widetilde{XY} &=&\widetilde{Y}\widetilde{X}.  \label{1.25d}
\end{eqnarray}

\textbf{eca-v }Let $I\in \bigwedge^{n}V.$ Then, for any $v\in V$ and $X\in
\bigwedge V:$%
\begin{equation}
I(v\wedge X)=(-1)^{n-1}v\lrcorner (IX).  \label{1.26}
\end{equation}
Eq.(\ref{1.26}) is sometimes called the \emph{duality identity} and since it
appears in several contexts in what follows we prove it.
\begin{proof}
By using eq.(\ref{1.23b}), $Iv=(-1)^{n-1}vI$ and $\widehat{I}=(-1)^{n}I$
where $v\in V$ and $I\in \bigwedge^{n}V$ and, eqs.(\ref{1.25c}) and (\ref
{1.23a}) we have
\begin{eqnarray*}
I(v\wedge X) &=&\frac{1}{2}(IvX+I\widehat{X}v)=\frac{1}{2}%
((-1)^{n-1}vIX+(-1)^{n}\widehat{I}\text{ }\widehat{X}v) \\
&=&(-1)^{n-1}\frac{1}{2}(vIX-\widehat{IX}v)=(-1)^{n-1}v\lrcorner
(IX).
\end{eqnarray*}
\end{proof}

\textbf{eca-vi }For any $X,Y,Z\in \bigwedge V:$%
\begin{eqnarray}
X\lrcorner (Y\lrcorner Z) &=&(X\wedge Y)\lrcorner Z,  \label{1.27a} \\
(X\llcorner Y)\llcorner Z &=&X\llcorner (Y\wedge Z).  \label{1.27b}
\end{eqnarray}

\begin{proof}
We prove only eq.(\ref{1.27a}). The proof of eq.(\ref{1.27b}) is
analogous and left to the reader.

Let $W\in \bigwedge V$. By using eq.(\ref{1.17a}) and eq.(\ref{1.9b}) we
have
\begin{eqnarray*}
(X\lrcorner (Y\lrcorner Z))\cdot W &=&(Y\lrcorner Z)\cdot (\widetilde{X}%
\wedge W)=Z\cdot ((\widetilde{Y}\wedge \widetilde{X})\wedge W) \\
&=&Z\cdot (\widetilde{(X\wedge Y)}\wedge W)=((X\wedge Y)\lrcorner Z)\cdot W.
\end{eqnarray*}
Hence, by the non-degeneracy of the euclidean scalar product, the first
statement follows.%
\end{proof}

To end, we call the readers attention to the fact that all Clifford algebra
associated to all possible euclidean structure $(V,G_{E})$ over the same
vector space $V$ are equivalent each to other, i.e., define the same
abstract Clifford algebra. Indeed all euclidean structures for $V$ are
isomorphic to the euclidean structure $(\mathbb{R}^{n},\bullet )$, where $%
\bullet $ is the canonical scalar product on $\mathbb{R}^{n}$. The
Clifford algebra associated to the euclidean structure
$(\mathbb{R}^{n},\bullet )$ is conveniently denoted
(\cite{33}\cite{34}) by $\mathbb{R}_{n}$.

\section{Conclusions}

The euclidean Clifford algebra\footnote{%
The classification of all euclidean algebras for arbitrary finite
dimensional space and their matrix representations can be found, e.g., in
\cite{30}.} $\mathcal{C\ell }(V,G_{E})$ introduced above will serve as our
basic calculational tool for the development of the theories of multivector
functions and multivector functionals that we develop in this series of
papers, and also for many applications that will be reported elsewhere. When
$\mathcal{C\ell }(V,G_{E})$ is used together with the concept of extensor
(to be introduced in paper II) we obtain a powerful formalism which permits
among other things an intrinsic presentation (i.e., without the use of
matrices) of the principal results of classical linear algebra theory. Also,
endowed $V$ with an arbitrary \emph{metric extensor} $g$ (of signature $%
s=p-q $ or $(p,q)$ as physicists like to say, with $p+q=n$) we can construct
a metric Clifford algebra $\mathcal{C\ell }(V,g)$ as a well-defined \emph{%
deformation} of the euclidean Clifford algebra $\mathcal{C\ell }(V,G_{E}),$
see paper IV.\bigskip

\textbf{Appendix A\medskip }

In the literature we can find several \emph{different} \emph{definitions}
(differing by numerical factors ($p!q!$ , $(p+q)!,$ $(p+q)!$/ $p!q!,$) for
the exterior product $X_{p}\wedge Y_{q}$ in terms of \emph{some} \emph{%
antisymmetrization} of the tensor product $X_{p}\otimes Y_{q}.$

Before continuing we recall that the tensor product of $t\in T^{p}V$ and $%
u\in T^{q}V,$ namely $t\otimes u\in T^{p+q}V,$ is defined by

(\textbf{ti}) for all $\alpha ,\beta \in T^{0}V\equiv
\mathbb{R}:\alpha \otimes \beta =\beta \otimes \alpha =\alpha
\beta $ (real product),

(\textbf{tii}) for all $\alpha \in \mathbb{R},u\in T^{q}V:\alpha
\otimes u=u\otimes \alpha =\alpha u$ (scalar multiplication of
$u$ by $\alpha $), and

(\textbf{tiii}) for all $t\in T^{p}V,u\in T^{q}V$ ($p,q\geq 1$) and $\omega
^{1},\ldots ,\omega ^{p},\omega ^{p+1},\ldots ,\omega ^{p+q}\in V^{*},$%
\begin{equation}
t\otimes u(\omega ^{1},\ldots ,\omega ^{p},\omega ^{p+1},\ldots ,\omega
^{p+q})=t(\omega ^{1},\ldots ,\omega ^{p})u(\omega ^{p+1},\ldots ,\omega
^{p+q})  \tag{A.1}
\end{equation}

Now, the exterior algebra $\bigwedge V$ is defined in the \emph{modern}
\emph{approach} to algebraic structures as the quotient $\bigwedge V=TV/I,$
where $TV=\sum\limits_{p=0}^{\infty }T^{p}V$ is the \emph{tensor algebra}
and $I$ is the \emph{bilateral ideal} generated by elements of the form $%
x\otimes x$. In this case, it is necessary to define the product of $%
X_{p}\in \bigwedge^{p}V$ and $Y_{q}\in \bigwedge^{q}V$ by
\begin{equation}
X_{p}\stackrel{qa}{\wedge }Y_{q}=\mathcal{A}(X_{p}\otimes Y_{q}),  \tag{A.2}
\end{equation}
instead of eq.(\ref{1.6bis}). This observation means that when reading books
with chapters on the theory of the exterior algebras or scientific papers,
it is necessary to take care and to be sure about which product has been
defined, for otherwise great confusion may arise. In particular for \emph{%
not }distinguishing $\wedge $ as defined in eq.(\ref{1.6bis}) from $%
\stackrel{qa}{\wedge }$ as defined by eq.(A.2) the following error appears
frequently. Let $X_{p}\in \bigwedge^{p}V,$ let $\{e_{i}\}$ be any basis of $%
V $ and $\{\varepsilon ^{i}\}$ its corresponding dual basis of $V^{*},$ and
consider the $p$ $1$-forms $\omega ^{1},\ldots ,\omega ^{p}\in V^{*}$. Then,
using the elementary expansions $\omega ^{1}(e_{j_{1}})\varepsilon
^{_{j_{1}}},\ldots $ etc., we have
\begin{eqnarray*}
X(\omega ^{1},\ldots ,\omega ^{p}) &=&X(\omega ^{1}(e_{j_{1}})\varepsilon
^{j_{1}},\ldots ,\omega ^{p}(e_{j_{p}})\varepsilon ^{j_{p}}) \\
&=&\omega ^{1}(e_{j_{1}})\ldots \omega ^{p}(e_{j_{p}})X(\varepsilon
^{j_{1}},\ldots ,\varepsilon ^{j_{p}}) \\
&=&X^{j_{1}\ldots j_{p}}e_{j_{1}}\otimes \ldots \otimes e_{j_{p}}(\omega
^{1},\ldots ,\omega ^{p}),
\end{eqnarray*}
i.e.,
\begin{equation}
X=X^{j_{1}\ldots j_{p}}e_{j_{1}}\otimes \ldots \otimes e_{j_{p}}.  \tag{A.3}
\end{equation}
The real numbers $X(\varepsilon ^{j_{1}},\ldots ,\varepsilon
^{j_{p}})=X^{j_{1}\ldots j_{p}}$ are called the $j_{1}\ldots j_{p}$-th
(contravariant) components of $X$ relative to the basis $\{e_{j_{1}}\otimes
\ldots \otimes e_{j_{p}}\}$ of $T^{p}V.$

Now, since $X\in \bigwedge^{p}V$ is a completly antisymmetric tensor it must
satisfy
\begin{equation}
\mathcal{A}X=X,  \tag{A.4}
\end{equation}
and using the definition of the operator $\mathcal{A}$ (see eq.(\ref{1.6novo}%
)) we get the identity
\begin{equation}
X^{j_{1}\ldots j_{p}}=\frac{1}{p!}\epsilon _{i_{1}\ldots i_{p}}^{j_{1}\ldots
j_{p}}X^{i_{1}\ldots i_{p}},  \tag{A.5}
\end{equation}
and of course the components $X^{j_{1}\ldots j_{p}}$ are antisymmetric in
all indices.

Using eq.(A.5) in eq.(A.3) we obtain,
\begin{equation}
X=\frac{1}{p!}\epsilon _{i_{1}\ldots i_{p}}^{j_{1}\ldots
j_{p}}X^{i_{1}\ldots i_{p}}e_{j_{1}}\otimes \ldots \otimes e_{j_{p}}.
\tag{A.6}
\end{equation}

Now, if we use the definition of the exterior product given by eq.(\ref
{1.6bis}), more exactly an particular case of eq.(\ref{1.6biss}), the
well-known combinatorial formula: $e_{i_{1}}\wedge \ldots \wedge
e_{i_{p}}=\epsilon _{i_{1}\ldots i_{p}}^{j_{1}\ldots j_{p}}e_{j_{1}}\otimes
\ldots \otimes e_{j_{p}},$ we see that eq.(A.6) can be written as
\begin{equation}
X=\frac{1}{p!}X^{i_{1}\ldots i_{p}}e_{i_{1}}\wedge \ldots \wedge e_{i_{p}}.
\tag{A.7}
\end{equation}

Eq.(A.7) is the expansion that has been used in this paper and in all the
others of this series.

Now, if we use the definition of the exterior product as given by eq.(A.2),
then by repeting the above calculations we get that $X$ can be witten as
\begin{equation}
X=X^{i_{1}\ldots i_{p}}e_{i_{1}}\stackrel{qa}{\wedge }\ldots \stackrel{qa}{%
\wedge }e_{i_{p}}.  \tag{A.8}
\end{equation}
To write $X=\dfrac{1}{p!}X^{i_{1}\ldots i_{p}}e_{i_{1}}\stackrel{qa}{\wedge }%
\ldots \stackrel{qa}{\wedge }e_{i_{p}}$ instead of eq.(A.8) is clearly wrong
if it is supposed that the meaning of $X(\varepsilon ^{j_{1}},\ldots
,\varepsilon ^{j_{p}})$ is that $X(\varepsilon ^{j_{1}},\ldots ,\varepsilon
^{j_{p}})=X^{j_{1}\ldots j_{p}}$ as in eq.(A.3). This confusion appears,
e.g., in \cite{36}.

\textbf{Acknowledgement}: V. V. Fern\'{a}ndez is grateful to FAPESP for a
posdoctoral fellowship. W. A. Rodrigues Jr. is grateful to CNPq for a senior
research fellowship (contract 251560/82-8) and to the Department of
Mathematics of the University of Liverpool for the hospitality. Authors are
also grateful to Doctors. P. Lounesto, I. Porteous, and J. Vaz, Jr. for
their interest on our research and for useful suggestions and discussions.

\end{document}